\documentstyle[12pt]{article}

\begin{document}

\mbox{} \hskip 10cm DF/IST-2.95

\mbox{} \hskip 10cm UATP-95/01

\mbox{} \hskip 10cm February 1995

\vskip 1cm

\begin{center}
BLACK HOLES IN THREE-DIMENSIONAL DILATON GRAVITY THEORIES \\
\vskip 1cm
{\bf Paulo M. S\'a} \\
\vskip 0.3cm
{\scriptsize Sector de F\'{\i}sica,
             Unidade de Ci\^encias Exactas e Humanas,
             Universidade do Algarve,} \\
{\scriptsize Campus de Gambelas,
             8000 Faro, Portugal}.
\vskip 0.6cm
{\bf Antares Kleber} \\
\vskip 0.3cm
{\scriptsize  Departamento de Astrof\'{\i}sica,
              Observat\' orio Nacional-CNPq,} \\
{\scriptsize  Rua General Jos\'e Cristino 77,
              20921 Rio de Janeiro, Brasil,} \\
\vskip 0.6cm
{\bf Jos\'e P. S. Lemos} \\
\vskip 0.3cm
{\scriptsize  Departamento de Astrof\'{\i}sica,
              Observat\' orio Nacional-CNPq,} \\
{\scriptsize  Rua General Jos\'e Cristino 77,
              20921 Rio de Janeiro, Brasil,} \\
{\scriptsize  \&} \\
{\scriptsize  Departamento de F\'{\i}sica,
              Instituto Superior T\'ecnico,} \\
{\scriptsize  Av. Rovisco Pais 1, 1096 Lisboa, Portugal.} \\
\end{center}

\bigskip

\begin{abstract}
\noindent
Three dimensional black holes in a generalized dilaton gravity action
theory are analysed.
The theory is specified by two fields,
the dilaton $\phi$ and the graviton $g_{\mu\nu}$,
and two parameters,
the cosmological constant $\lambda$ and the Brans-Dicke parameter $\omega$.
It contains seven different cases,
of which one distinguishes as special cases,
string theory,
general re\-la\-tivity
and a theory equivalent to four dimensional general relativity
with one Killing vector.
We study the causal structure and geodesic motion of null and
timelike particles in the black hole geometries and find the ADM
masses of the different solutions.
\end{abstract}

\newpage

\noindent
{\bf 1. Introduction}

\vskip 3mm

Black holes are fundamental objects in classical
theories of gravitation and,
through Hawking radiation,
are of central importance for the understanding of the physical
processes that occur at a quantum gravity scale.

Black holes were first predicted within four dimensional General
Relati\-vi\-ty as structures emerging out of complete gravitational
collapse of massive objects \cite{whee}.
More recently,
they have also appeared as exact solutions of
se\-ve\-ral gravity theories in two, three, four and higher dimensions.
String theory is an example of a theory that has provided a great variety
of black hole solutions as well as extended black objects such as
black  strings and black membranes \cite{horo}.
Most of these solutions are exact solutions of the low energy action
of heterotic string theory.

The study of black holes in dimensions lower than four has proved
fruitful in  two aspects:
(i) a better understanding of the physical features
(such as temperature, entropy, radiated flux) in a black hole geometry,
(ii) finding the corresponding exact conformal field theory.
In relation with the latter,
it has been shown  that the two dimensional black hole of string theory
is a solution of the full classical action corresponding to an
exact conformal field theory \cite{witt}
and that the three dimensional black string can also be recovered
from such an exact conformal field theory \cite{horn}.
This has not yet ocurred for higher dimensions.

Surprisingly,
the only three dimensional black hole in string theory is
obtained by taking the product of the two dimensional black hole
with $S^1$ \cite{horo}.
On the other hand,
three dimensional General Relativity with a negative cosmological constant
contains a black hole solution with constant curvature \cite{bana}.
This black hole has mass, temperature and entropy
and in addition can also have charge and  angular momentum.
It has therefore many features similar to the Kerr-Newman black holes
in four dimensional General Relativity.

Our aim in this article is to study three dimensional (3D) static
black holes in a theory that contains as particular cases
General Relativity and string theory.
For that  purpose we consider a generalized  dilaton
gravity action (i.e. a Brans-Dicke action) in 3D.
This theory is specified by two fields,
the dilaton $\phi$ and the graviton $g_{\mu\nu}$,
and two parameters,
the cosmological constant $\lambda$ and the Brans-Dicke parameter $\omega$.

The plan of this article is the following.
Section 2 sets up the action and the equations of motion.
We specify the form of the metric.
The ansatz we use is not the most general 3D static metric yet
it can give an important cross section of the black holes
in different $\omega$ theories.
We start section 3 by finding the general solution of the equations of
motion and write the corresponding Kretschmann scalar which
signals the presence of singularities.
In section 4 we use an extension of the formalism of Regge and Teitelboim
to derive the masses of the black holes.
In section 5 we study the seven different cases that appear
naturally from the solutions.
We work out in detail the causal structure
and the geodesic motion of null and timelike particles
for each case separately.
Thus section 5 is divided in seven subsections.
In section 6 we present the concluding remarks.

\vskip 1cm

\noindent
{\bf 2. The Equations}

\vskip 3mm

We are going to work with the following action,
\begin{equation}
S=\frac{1}{2\pi} \int d^3x \sqrt{-g} e^{-2\phi}
  \left( R - 4 \omega \left( \partial \phi \right)^2
           + 4 \lambda^2
  \right),                                   \label{eq:1}
\end{equation}
where
$g$ is the determinant of the 3D metric,
$R$ is the curvature scalar,
$\phi$ is a scalar field,
$\lambda$ is a constant
and $\omega$ is a parameter.
Equation (\ref{eq:1}) is a Brans-Dicke action in three-dimensions.

Varying this equation with respect to $\phi$ and $g^{ab}$
one gets the following equations,
\begin{eqnarray}
   & &  R -4\omega D_cD^c\phi
          +4\omega D_c\phi D^c\phi
          +4\lambda^2 =0,      \label{eq:3} \\
   & &   \frac12 G_{ab}
       -2\left(\omega+1\right) D_a\phi D_b\phi
       +D_aD_b\phi
       -g_{ab} D_cD^c\phi   \nonumber \\
   & &  \hskip 4cm
       +\left(\omega+2\right) g_{ab} D_c\phi D^c\phi
       -g_{ab}\lambda^2=0,
                               \label{eq:2}
\end{eqnarray}
where $G_{ab}=R_{ab}-\frac12 g_{ab} R$ is the Einstein tensor and
$D$ represents the covariant derivative.

We want to obtain static black hole solutions.
The most general static metric with a Killing vector
$\partial/\partial\varphi$ with closed orbits in three dimensions
can be written in the Schwarzschild gauge as \cite{melv}:
\begin{equation}
     ds^2 = - e^{2\nu(r)} dt^2 + e^{2\mu(r)}dr^2 + r^2 d\varphi^2,
              \quad\quad 0\leq\varphi\leq 2\pi.
                               \label{eq:4a}
\end{equation}

Now,
each different $\omega$ can be viewed as yielding a different
dilaton gravity theory.
For instance,
for $\omega=-1$ one gets the simplest low energy string action \cite{call},
for $\omega=0$ one gets a theory related to
General Relativity with one Killing vector \cite{lemo},
for $\omega=\infty$ one obtains three dimensional General Relativity
\cite{bana}.
Each different $\omega$ has a very rich and non-trivial structure
of solutions which could be considered on its own.
Here we do not attempt to study the whole set of solutions of
the different theories.
Rather,
we drastically restrict the general metric (\ref{eq:4a})
in order to probe, with particular solutions, each theory.
In this way one can compare different black hole solutions in
different theories.
Our choice is to work in a particular Schwarzschild gauge,
$\mu(r)=-\nu(r)$,
where the metric is written as:
\begin{equation}
     ds^2 = - e^{2\nu(r)} dt^2 + e^{-2\nu(r)}dr^2 + r^2 d\varphi^2.
                               \label{eq:4}
\end{equation}
Inserting this metric in equations (\ref{eq:3}) and (\ref{eq:2})
one obtains the following set of four differential equations:
\begin{eqnarray}
  & &  \omega \phi_{,rr}
      + 2 \omega \phi_{,r} \nu_{,r}
      + \omega \frac{\phi_{,r}}{r}
      - \omega (\phi_{,r})^2
      + \frac{\nu_{,r}}{r}
      + \frac{\nu_{,rr}}{2}
      + (\nu_{,r})^2
      - \lambda^2 e^{-2\nu}=0,
                                    \label{eq:5}  \\
  & &   \phi_{,rr}
      + \phi_{,r} \nu_{,r}
      + \frac{\phi_{,r}}{r}
      - (\omega+2) (\phi_{,r})^2
      - \frac{\nu_{,r}}{2r}
      + \lambda^2 e^{-2\nu}=0,
                                    \label{eq:6}  \\
  & &  - \phi_{,r} \nu_{,r}
       - \frac{\phi_{,r}}{r}
       - \omega (\phi_{,r})^2
       + \frac{\nu_{,r}}{2r}
       - \lambda^2 e^{-2\nu}=0,
                                    \label{eq:7}  \\
  & &    \phi_{,rr}
       + 2 \phi_{,r} \nu_{,r}
       - (\omega+2) (\phi_{,r})^2
       - \frac{\nu_{,rr}}{2}
       - (\nu_{,r})^2
       + \lambda^2 e^{-2\nu}=0,
                                    \label{eq:8}
\end{eqnarray}
where ${}_{,r}$ denotes a derivative with respect to $r$.

\vskip 1cm

\noindent
{\bf 3. The general solution}

\vskip 3mm

Let us first consider the case $\omega\neq -1$.
Summing up equations (\ref{eq:6}) and (\ref{eq:7}) one obtains
$\phi_{,rr}=2(\omega+1)(\phi_{,r})^2$,
yielding for the dilaton field the following solution:
\begin{equation}
\phi = -\frac{1}{2(\omega+1)}
        \ln \left[ 2(\omega+1)r+A_1 \right] + A_2,
                                   \label{eq:9}
\end{equation}
where $A_1$ and $A_2$ are constants of integration.
One can, without loss of generality, choose $A_1=0$.
Then, eq.~(\ref{eq:9}) can be written as:
\begin{equation}
  e^{-2\phi}= A (ar)^{\frac{1}{\omega+1}},
                                    \label{eq:9a}
\end{equation}
where $a$ is a constant (see below).
The dimensionless constant $A$ can be viewed
as a normalization to the action (\ref{eq:1}).
Since it has no influence in our calculations,
apart a possible redefiniton of the mass (see section~4),
we set $A=1$.

Inserting the solution (\ref{eq:9a}) in equations
(\ref{eq:5})-(\ref{eq:8}), we obtain for the metric:
\begin{eqnarray}
 ds^2 &=& -\left[
            (ar)^2 - \frac{b}{(ar)^{\frac{1}{\omega+1}}}
            \right] dt^2
           + \frac{dr^2}{(ar)^2 - \frac{b}{(ar)^{\frac{1}{\omega+1}}}}
           + r^2 d\varphi^2,                      \nonumber \\
      & &  \hskip 5cm \omega \neq -2,-\frac32,-1,
                                             \label{eq:10}  \\
 ds^2 &=& -\left( br - r^2 \right) dt^2
           + \frac{dr^2}{br - r^2}
           + r^2 d\varphi^2,                      \nonumber \\
      & &  \hskip 5cm \omega =-2,
                                             \label{eq:11}  \\
 ds^2 &=&    4\lambda^2 r^2 \ln (br) dt^2
           - \frac{dr^2}{4\lambda^2 r^2 \ln (br)}
           + r^2 d\varphi^2,                      \nonumber \\
      & &  \hskip 5cm \omega=-\frac32,
                                             \label{eq:12}
\end{eqnarray}
where $b$ is a constant of integration.
For the cases covered by equations (\ref{eq:9a}) and (\ref{eq:10})
$a$ is defined as:
\begin{equation}
 a = \frac{2(\omega+1) | \lambda|}{\sqrt{(\omega+2)(2\omega+3)}}.
\end{equation}
For the other cases
(covered by equations (\ref{eq:9a}), (\ref{eq:11}) and (\ref{eq:12}))
we set $a=1$.
For $\omega=-2$ equations~(\ref{eq:5})-(\ref{eq:8}) imply $\lambda=0$.
Since we are interested in solutions with horizons,
we will take the constant $b$ to be positive.
As it will be shown below, this constant
is related with the mass of the solutions.

For $\omega=-1$, it follows from equations
(\ref{eq:5})-(\ref{eq:8}) that $\lambda=0$, $\nu=C_1$,
$\phi=C_2$, where $C_1$ and $C_2$ are constants of integration.

An important quantity,
which signals the appearence of a curvature singularity,
is the Kretschmann scalar:
\begin{eqnarray}
  R_{abcd}R^{abcd} &=&
       12a^4
     + \frac{4\omega}{(\omega+1)^2}
       \frac{ba^4}{(ar)^{\frac{2\omega+3}{\omega+1}}}
     + \frac{3\omega^2 + 8\omega +6}{(\omega+1)^4}
       \frac{b^2a^4}{(ar)^{\frac{4\omega+6}{\omega+1}}},
                         \nonumber \\
         & &     \hskip 6.3cm \omega \neq -2,-\frac32,-1,
                                             \label{eq:13}  \\
  R_{abcd}R^{abcd} &=&
       12 - \frac{8b}{r} + \frac{2b^2}{r^2},
              \hskip 3.6cm \omega =-2,
                                             \label{eq:14}  \\
  R_{abcd}R^{abcd} &=&
       16\lambda^4 \left[
           11 + 20 \ln (br) + 12 \ln^2 (br)
                   \right],
             \hskip 0.3cm \omega=-\frac32.
                                             \label{eq:15}
\end{eqnarray}
An inspection of the Kretschmann scalar reveals that for
$-\frac32<\omega<-1$ the curvature singularity is located at
$r=+\infty$ and for $-\frac32>\omega>-1$ the singularity is
at $r=0$.
For $\omega=-\frac32$ both $r=0$ and $r=+\infty$ are singular.
For $\omega=\infty$ spacetime has no
curvature singularities.

\vskip 1cm

\noindent
{\bf 4. The ADM mass of the solutions}

\vskip 3mm

The solutions (\ref{eq:10})-(\ref{eq:12}) allow a meaningful
definition of mass.
Here we apply the formalism  of Regge and Teitelboim \cite{regg}
(see also \cite{lemo}).

We write the metric (\ref{eq:4}) in the canonical form
\begin{equation}
     ds^2 = - (N^0)^2 dt^2
            + \frac{dr^2}{f^2}
            + r^2 d\varphi^2,
                               \label{eq:16}
\end{equation}
where $N^0(r)$ is the lapse function and $f(r)$
is a well defined function.
Then,
the Hamiltonian form of the action can be written as:
\begin{eqnarray}
 & & S = \Delta t \int dr N
        \left[ 4f^2 (r \phi_{,r} e^{-2\phi})_{,r}
      + (2r\phi_{,r} -1) (f^2)_{,r} e^{-2\phi} \right. \nonumber \\
 & &  \ \ \ \ \ \ \ \ \ \ \ \
      - \left. 4\omega r f^2 (\phi_{,r})^2 e^{-2\phi}
      + 4\lambda^2 r e^{-2\phi} \right]
      + B.
                               \label{eq:17}
\end{eqnarray}
$N=\frac{N^0}{f}$ is a Lagrange multiplier imposing a
constraint on the action and $B$ is a surface term.
Upon varying the action with respect to $\delta f$ and $\delta \phi$
one picks up aditional surface terms.
Indeed,
\begin{eqnarray}
\delta S &=& - \Delta t N \left[
        (1-2r\phi_{,r}) e^{-2\phi} \delta f^2
      - 4f^2 \delta (r \phi_{,r} e^{-2\phi})
                          \right. \nonumber \\
         & & \ \ \ \  \left.
      + 8\omega r f^2 \phi_{,r} e^{-2\phi} \delta \phi
      + 2 r (f^2)_{,r} e^{-2\phi} \delta \phi
                          \right] + \delta B \nonumber \\
         &+& (\mbox{terms vanishing when the
                    equations of motion hold}).
                               \label{eq:18}
\end{eqnarray}
In order that the Hamilton's equations are satisfied,
the surface term $B$ has to be adjusted so that it cancels the first
four terms on the right hand side of equation (\ref{eq:18}).
Taking in account that in the gauge of eq.~(\ref{eq:16})
$\delta \phi=0$, we obtain:
\begin{equation}
  \delta B = -\Delta t N \left[ (2r\phi_{,r}-1)
                  e^{-2\phi} \delta f^2 \right].
                               \label{eq:19}
\end{equation}
The term in square brackets gives the mass,
since it is the term conjugate to $N$.
Inserting the solutions (\ref{eq:9a})-(\ref{eq:12}),
one obtains for the mass:
\begin{eqnarray}
 M &=& \frac{\omega + 2}{\omega +1} b,
       \hskip 2.6cm \omega\neq -\frac32,-1, \label{eq:19a} \\
 M &=& -4\lambda^2 \ln (b),
       \hskip 2cm \omega = -\frac32.        \label{eq:19b}
\end{eqnarray}

Note that if we have not set $A=1$ in eq.~(\ref{eq:9a})
all the masses $M$ would be rescaled by the factor $A$ itself.

\vskip 1cm

\noindent
{\bf 5. Causal structure and geodesic motion for the seven different solutions}

\vskip 0.3cm

Given a black hole spacetime it is of great interest to study its
geodesics,
since in turn they can shed some light on the nature of the spacetime itself.
The equations governing the geodesics can be derived from the Lagrangian
\begin{equation}
{\cal L} = \frac12 g_{ab}\frac{dx^a}{d\tau} \frac{dx^b}{d\tau},
                               \label{eq:20}
\end{equation}
where $\tau$ is some affine parameter along the geodesic.
For timelike geodesics $\tau$ can be identified with the proper time
of the particle moving along the geodesic.
Inserting the metric (\ref{eq:4}) one obtains
\begin{equation}
  \left( \frac{dr}{d\tau}\right)^2 + V(r) = E,
                               \label{eq:21}
\end{equation}
where
\begin{equation}
 V(r) = e^{2\nu} \left( \alpha +\frac{L^2}{r^2} \right)
                               \label{eq:22}
\end{equation}
is the potencial, $L$ is the angular momentum, $\alpha=1$ for
timelike geodesics and $\alpha=0$ for null geodesics.
$E$ is a constant, which cannot be interpreted as the energy of the
particle at infinity.
Inspection of eq.~(\ref{eq:21}) will allow in a simple manner
to analyse geodesic motion.
Since inside the horizon $r$ and $t$ change roles
($r$ becomes the temporal coordinate),
equations (\ref{eq:21}) and (\ref{eq:22}) will be used to identify
closed timelike curves.

It is also important to study the causal structure of each spacetime.
In order to do so we define the advanced and retarded
null coordinates,
\begin{equation}
  u=t-r_*,    \quad\quad
  v=t+r_*.
                              \label{eq:23}
\end{equation}
Then, the metric (\ref{eq:4}) becomes:
\begin{equation}
  ds^2 = - e^{2\nu(r)} dudv + r^2 d\varphi^2,
                              \label{eq:24}
\end{equation}
where $r_*$ is the ``tortoise'' coordinate
\begin{equation}
    r_*=\int  e^{-2\nu(r)} dr.
                              \label{eq:25}
\end{equation}
The maximal analytical extension is performed by
introducing Kruskal coordinates $U$ and $V$ (see below).

In general,
the integral in eq.~(\ref{eq:25}) does not have an
analytical expression for the solutions given by
eqs.~(\ref{eq:10})-(\ref{eq:12}).
Moreover,
the maximal analytical extension depends critically
on the values of $\omega$.
There are seven cases which have to be treated separately:
$(-\infty<\omega<-\frac32)-\{-2\}$,
$\omega=-2$,
$\omega=-\frac32$,
$-\frac32<\omega<-1$,
$\omega=-1$,
$-1<\omega<+\infty$
and
$\omega=\pm\infty$.

\vskip 1cm

\noindent
{\bf 5.1  $(-\infty<\omega<-\frac32)-\{-2\}$}

\vskip 0.3cm

Within this range of values of $\omega$ there is no general
analytical solution for the integral in eq.~(\ref{eq:25}).
Thus,
we are going to analyse $\omega=-3$,
which is a typical case.
As it will be shown this solution corresponds to a black hole.

The dilaton and the metric are
\begin{eqnarray}
  & &  e^{-2\phi}=\frac{1}{\sqrt{ar}},
                            \label{eq:26} \\
  & &  ds^2 = - \left[ (ar)^2 - \sqrt{ar} \right] dt^2
              + \frac{dr^2}{(ar)^2 - \sqrt{ar}}
              + r^2 d\varphi^2,
                            \label{eq:27}
\end{eqnarray}
with $a=-\frac{4\sqrt3 |\lambda|}{3}$.

Since for different (positive) values of $b$ the causal
structure and the cha\-racter of geodesic motion do not change,
we have taken here and will take throughout this paper $b=1$,
except when analysing the mass of the solutions.

To remove the coordinate singularity at the horizon ($ar=1$)
we introduce the Kruskal coordinates
\begin{equation}
  U = \mp\frac{4}{3a}
         \exp \left( -\frac{3a}{4}u \right), \quad\quad
  V =    \frac{4}{3a}
         \exp \left(  \frac{3a}{4}v \right),
                            \label{eq:28}
\end{equation}
where $\mp$ correspond, respectively,
to region I ($ar>1$) and to region II ($0<ar<1$).
Now the metric takes the form
\begin{equation}
    ds^2 = - \sqrt{ar}
             \left( ar+\sqrt{ar}+1 \right)^{3/2}
             e^{\sqrt{3} \arctan \frac{\sqrt{3ar}}{2+\sqrt{ar}}}
             dUdV
           + r^2 d\varphi^2,
                            \label{eq:29}
\end{equation}
where $r=r(U,V)$.
To regions I and II there correspond regions I' and II',
which are simply a replic of the former.
The true spacetime singularity is located at $ar=0$.
The timelike infinity can be properly defined at $ar\rightarrow \infty$.
See figure~1 for the Penrose diagram.

The potential (\ref{eq:22}),
\begin{equation}
 V(r)=\left[ (ar)^2 - \sqrt{ar} \right]
      \left( \alpha + \frac{L^2}{r^2} \right),
                            \label{eq:30}
\end{equation}
indicates that, in region I, there are no bounded orbits
for timelike and null geodesic motion. Particles moving along
null geodesics (for which $E<a^2 L^2\neq 0$)
and timelike geodesics will fall inside the black hole.
Null geodesics for which $L=0$ or $E>a^2 L^2\neq 0$ reach
spatial infinity for a infinite value of its affine parameter $\tau$.
Inside the horizon, coordinates $r$ and $t$ change roles;
in particular, $r$ becomes a temporal coordinate.
Equations (\ref{eq:21}) and (\ref{eq:30}) then show that there are closed
timelike curves for certain values of $L$ and $E$.

Evaluating eq.~(\ref{eq:19a}) for the mass we obtain for the
$\omega=-3$ black hole:
\begin{equation}
  M = \frac{b}{2}.           \label{eq:31}
\end{equation}
 From equation (\ref{eq:19a}) it follows that the mass
is positive for the black holes in the range of values of
$\omega$ from $-\infty$ to $-2$,
and negative for $-2<\omega<-\frac32$.

\vskip 1cm

\noindent
{\bf 5.2  $\omega=-2$}

\vskip 3mm

This solution corresponds to a naked singularity.

The dilaton and the metric are
\begin{eqnarray}
  & &  e^{-2\phi}=\frac{1}{r},
                            \label{eq:32} \\
  & &  ds^2 = - \left( -r^2 + r \right) dt^2
              + \frac{dr^2}{-r^2 +r}
              + r^2 d\varphi^2.
                            \label{eq:33}
\end{eqnarray}
There is a coordinate singularity at $r=1$.
To remove it we introduce the Kruskal coordinates
\begin{equation}
  U = \mp 2 \exp \left(-\frac12 u \right), \quad\quad
  V =     2 \exp \left( \frac12 v \right),
                            \label{eq:34}
\end{equation}
where $\mp$ correspond, respectively,
to region I ($0<r<1$) and to region II ($r>1$).
Now the metric reads:
\begin{equation}
    ds^2 = - r^2 dUdV
           + r^2 d\varphi^2.
                            \label{eq:35}
\end{equation}

To regions I and II there correspond regions
I' and II', which are simply a replic of the former.
The singularity at $r=0$ is a naked null singularity.
See figure~2 for the Penrose diagram.

Analysing the potential (\ref{eq:22})
\begin{equation}
 V(r)=\left( -r^2 + r \right)
      \left( \alpha + \frac{L^2}{r^2} \right),
                            \label{eq:36}
\end{equation}
one concludes that there are bounded timelike orbits for values of
the angular momentum in the range $0<L^2<1/27$.
For null geodesic motion there are no bounded orbits.
Particles with $L \neq 0$ moving along null and timelike geodesics
will not hit the singularity; those that not describe bounded orbits
will fall inside region II. In this region, both null and timelike geodesics,
can be extended up to arbitrary large values of the affine parameter $\tau$.
For both timelike and null geodesic motion the naked
singularity can be reached only for $L=0$.

It is straightforward to conclude from eq.~(\ref{eq:19a}) that
the mass equals zero.

\vskip 1cm

\noindent
{\bf 5.3  $\omega=-\frac32$}

\vskip 3mm

This solution corresponds to a spacetime whose whole frontier
is singular.

The dilaton and the metric are
\begin{eqnarray}
  & &  e^{-2\phi}=\frac{1}{r^2},
                            \label{eq:37} \\
  & &  ds^2 =   4\lambda^2 r^2 \ln r dt^2
              - \frac{dr^2}{4\lambda^2 r^2 \ln r}
              + r^2 d\varphi^2.
                            \label{eq:38}
\end{eqnarray}
To remove the coordinate singularity at $r=1$ (horizon)
we introduce the Kruskal coordinates
\begin{equation}
  U = \mp\frac{1}{2\lambda^2}
         \exp \left( -2\lambda^2 u \right), \quad\quad
  V =    \frac{1}{2\lambda^2}
         \exp \left(  2\lambda^2 v \right),
                            \label{eq:39}
\end{equation}
where $\mp$ correspond, respectively,
to region I ($0<r<1$) and to region II ($r>1$).
Now the metric takes the form
\begin{equation}
    ds^2 = - 4\lambda^2 r^2
             \exp \left[ -\sum_{k=1}^{\infty}
                           \frac{(-1)^k(\ln r)^k}{k k!}
                  \right]
             dUdV
           + r^2 d\varphi^2.
                            \label{eq:40}
\end{equation}
To regions I and II there correspond regions
I' and II', which are simply a replic of the former.
The whole frontier of spacetime is singular;
indeed, the Kretschmann scalar (see eq.~(\ref{eq:15})) diverges for
$r=0$ and $r=+\infty$.
That $r=+\infty$ is spacelike can be obtained by calculating
the limit $r\rightarrow +\infty$ of $UV$,
\begin{eqnarray}
   UV &=& \frac{1}{4\lambda^2}
         \exp \left[ \ln | \ln r |
                    +\sum_{k=1}^{\infty}
                           \frac{(-1)^k(\ln r)^k}{k k!}
              \right],
                            \nonumber      \\
      &=& \frac{1}{4\lambda^2}
          \exp \left[ \int_{0}^{1/r}
                      \frac{dy}{\ln y}
               \right], \quad\quad (r>1)
                            \label{eq:41}  \\
      &=& -\frac{1}{4\lambda^2}
          \exp \left[ \int_{-\infty}^{\ln(1/r)}
                      \frac{e^{y} dy}{y}
               \right], \quad\quad  (0<r<1),
                            \label{eq:42}
\end{eqnarray}
which is $\frac{1}{4\lambda^2}$, i.e.\ a horizontal hyperbola
in the Kruskal diagram or a horizontal line in the Penrose diagram.
Making a similar calculation for $r=0$, we obtain that $UV=-\infty$,
which corresponds to 45 degree lines in the Penrose diagram (see figure~3).

The potential (\ref{eq:22})
\begin{equation}
 V(r)=-4\lambda^2 r^2 \ln r
        \left( \alpha + \frac{L^2}{r^2} \right),
                            \label{eq:43}
\end{equation}
indicates that, in region I, there are bounded orbits
for timelike geodesic motion for certain values of $E$ and $L$.
Timelike and null geodesics for which $L\neq0$ do not reach the
naked singularity at $r=0$.
Those that not describe bounded orbits will enter region II,
where they can be extended up to
infinite values of the affine parameter $\tau$.
The analysis of radial ($L=0$) timelike and null
geodesic motion indicates that singularity at $r=0$ can be reached
for finite values of the affine parameter.

For the mass of this solution we have that
$M=-4\lambda^2 \ln (b)$ (see eq.~(\ref{eq:19b})).
Depending on the value of the constant of integration $b$,
the mass can be positive, negative or zero.

\vskip 1cm

\noindent
{\bf 5.4  $-\frac32<\omega<-1$}

\vskip 3mm

Within this range of values of $\omega$ there is no general
analytical solution for the integral in eq.~(\ref{eq:25}).
Thus,
we are going to analyse here the case $\omega=-\frac43$.
As it will be shown this solution corresponds to a black hole.

The dilaton and the metric are
\begin{eqnarray}
  & &  e^{-2\phi}=\frac{1}{(ar)^3},
                            \label{eq:44a} \\
  & &  ds^2 = - \left[ (ar)^2 - (ar)^3 \right] dt^2
              + \frac{dr^2}{(ar)^2 - (ar)^3}
              + r^2 d\varphi^2,
                            \label{eq:44}
\end{eqnarray}
with $a=-\sqrt2 |\lambda|$.
The coordinate singularity at the horizon ($r=1$) is removed by
the introduction of the following Kruskal coordinates:
\begin{equation}
  U = \mp\frac{2}{a}
         \exp \left(  \frac{a}{2}u \right), \quad\quad
  V =    \frac{2}{a}
         \exp \left( -\frac{a}{2}v \right),
                            \label{eq:45}
\end{equation}
where $\mp$ correspond, respectively,
to region I ($0<ar<1$) and to region II ($ar>1$).
In these coordinates the metric takes the form
\begin{equation}
    ds^2 = - (ar)^3
             e^{-\frac{1}{ar}}
             dUdV
           + r^2 d\varphi^2.
                            \label{eq:46}
\end{equation}
The physical singularity is located inside the horizon at $ar=+\infty$,
while at $ar=0$ there is a topological singularity.
For this reason the spacetime manifold cannot be extended for
negative values of $r$, contrary to the two-dimensional case \cite{lesa}.
For the Penrose diagram see figure~4.

Let us now consider geodesic motion.
An analysis of the potential (\ref{eq:22})
\begin{equation}
 V(r)=\left[ (ar)^2 - (ar)^3 \right]
      \left( \alpha + \frac{L^2}{r^2} \right),
                            \label{eq:47}
\end{equation}
indicates that, in region I, there are bounded orbits
for timelike geodesic motion if $0<a^2 L^2<\frac13$.
For null geodesic motion there are no bounded orbits.
Both timelike and null geodesics can be extended up to the conical
singularity at $r=0$ for a finite value of the affine parameter $\tau$.
Inside the horizon it will take a finite proper time for a observer
moving along a timelike geodesic to fall into the singularity.
Null geodesics can be extended up to infinite values of the affine parameter.
Thus,
except for those values of $L$ and $E$ corresponding to bounded orbits,
all timelike geodesic observers will fall into a singularity
(physical or conical) in a finite proper time.

In this range of values of $\omega$ we have to distinguish between two
cases: $-\frac32<\omega \leq -\frac43$
(including the case we have been considering)
and $-\frac43<\omega<-1$.
Although both cases have the same causal structure,
they differ in what concerns geodesic motion, the difference being
the fact that for the latter case there are
no bounded orbits in region I.
Thus,
timelike geodesic motion will end inevitably in a singularity
after a finite amount of proper time have been elapsed.

For $\omega=-\frac43$ the mass is given by (see eq.~(\ref{eq:19a})):
\begin{equation}
  M = -2b.   \label{eq:48}
\end{equation}
For the range of values of $\omega$ we are considering in this subsection,
the mass is always negative.

\vskip 1cm

\noindent
{\bf 5.5  $\omega=-1$}

\vskip 3mm

This case is interesting since $\omega=-1$ corresponds to the simplest
low energy action in string theory.

Here,
the dilaton field is constant and the metric reads
\begin{equation}
  ds^2 = - dt^2 + dr^2 + r^2 d\varphi^2,
                            \label{eq:49}
\end{equation}
which is the metric of three-dimensional Minkowski spacetime.
Thus,
there are no black hole solutions.
This is a known result \cite{horo,horn};
there are no static axisymmetric black holes in three dimensional
string theory apart a very special one.
This special black hole is trivially obtained through the product
of the two dimensional black hole \cite{witt,mand} with $S^1$.
It can also be obtained through eq.~(\ref{eq:10}),
if one makes the following singular coordinate transformation:
\begin{equation}
   r \rightarrow \bar{r}=\frac{1}{a} \ln (ar).
\end{equation}
Then,
the metric (\ref{eq:10}) becomes:
\begin{equation}
    ds^2 = - \left( 1 - b e^{-2 | \lambda | \bar{r}} \right) dt^2
           + \frac{d\bar{r}^2}{1 - b e^{-2 | \lambda | \bar{r}}}
           + d\bar{\varphi}^2,
\end{equation}
where $d\bar{\varphi}^2 \equiv \frac{1}{a^2}d\varphi^2$.
The dilaton field in this case is given by
$e^{-2\phi}=e^{2|\lambda|r}$.

\vskip 1cm

\noindent
{\bf 5.6  $-1<\omega<+\infty$}

\vskip 3mm

Within this range of values of $\omega$ there is no general
analytical solution for the integral in eq.~(\ref{eq:25}).
Thus,
we are going to analyse $\omega=0$, which is a typical case.

The black hole solution obtained in this case was first
discussed in ref.~\cite{lemo},
where the three-dimensional gravity theory
was obtained through dimensional reduction from four dimensional
General Relativity with one Killing vector field.

The dilaton and the metric are
\begin{eqnarray}
  & &  e^{-2\phi}= ar,
                            \label{eq:50} \\
  & &  ds^2 = - \left[ (ar)^2 - \frac{1}{ar} \right] dt^2
              + \frac{dr^2}{(ar)^2 - \frac{1}{ar}}
              + r^2 d\varphi^2,
                            \label{eq:51}
\end{eqnarray}
with $a=\frac{\sqrt6 |\lambda|}{3}$.
Introducing Kruskal coordinates
\begin{equation}
  U = \mp\frac{2}{3a}
         \exp \left( -\frac{3a}{2}u \right), \quad\quad
  V =    \frac{2}{3a}
         \exp \left(  \frac{3a}{2}v \right),
                            \label{eq:52}
\end{equation}
where $\mp$ correspond, respectively,
to region I ($ar>1$) and to region II ($0<ar<1$),
the coordinate singularity at $ar=1$ is removed,
yielding a metric which takes the form
\begin{equation}
    ds^2 = - \frac{\left[ (ar)^2+ar+1 \right]^{3/2}}{ar}
             e^{-\sqrt{3} \arctan \frac{2ar+1}{\sqrt{3}}}
             dUdV
           + r^2 d\varphi^2.
                            \label{eq:53}
\end{equation}
The true spacetime singularity is located at $ar=0$ and
the infinity $r=+\infty$ is timelike.
Thus,
the causal structure of this black hole solution
is similar to the one discussed in section~3.1
(see figure~1 for the Penrose diagram).

 From the potential (\ref{eq:22})
\begin{equation}
 V(r)=\left[ (ar)^2 - \frac{1}{ar} \right]
      \left( \alpha + \frac{L^2}{r^2} \right),
                            \label{eq:54}
\end{equation}
it follows that there are no bounded null or timelike geodesic motion.
An observer moving in region I along a timelike geodesic
will inevitably cross the horizon and enter the black hole,
where an finite proper time is required for him to hit the singularity.
Null geodesics for which $L=0$ or $E>a^2 L^2 \neq 0$ can be,
in region I,
extended up to arbitrary large values of the affine parameter $\tau$.
In region II,
null geodesics will reach the singularity for finite values of $\tau$.

Evaluating eq.~(\ref{eq:19b}) for the mass we obtain
\begin{equation}
  M = 2 b.                 \label{eq:55}
\end{equation}
The mass is positive for any of the values of $\omega$
in consideration in this subsection.

\vskip 1cm

\noindent
{\bf 5.7 $\omega=\mp\infty$}

\vskip 3mm

For these values of the parameter $\omega$
the dilaton field is constant and the metric is given by
\begin{equation}
  ds^2 = - \left[ (ar)^2 - b \right] dt^2
              + \frac{dr^2}{(ar)^2 -b}
              + r^2 d\varphi^2,
  \quad\quad a=\sqrt2 |\lambda|,
                            \label{eq:56}
\end{equation}
corresponding to the three-dimensional anti-de Sitter spacetime.
Upon identifying certain points of the anti-de Sitter
space by means of a discrete subgroup of its isometry
group $SO(2,2)$ a black hole arises \cite{bana2}.
The causal structure of the black hole solution
is similar to the one exhibited in figure~1 with the difference
that the surface $r=0$ is not a curvature singularity but, rather,
a singularity in the causal structure.
For a detailed discussion of this black hole see ref.~\cite{bana,bana2}.
The geodesics are analysed in ref.~\cite{fari}.

\vskip 1cm

\noindent
{\bf 6. Conclusions}

\vskip 3mm

We have found static black hole solutions in a generalized 3D dilaton
gravity specified by the Brans-Dicke parameter $\omega$.
Although we have worked in a particular Schwarzschild gauge metric,
one knows that there are two theories for which
the corresponding black hole is unique.
They are General Relativity
$\left( \omega\rightarrow\infty\right)$ \cite{bana2},
and string theory \cite{horo}
($\omega=-1$).
However,
for $\omega=0$ we suspect that there is a whole
class of different black holes.
This is because $\omega=0$ is equivalent to four dimensional
General Relativity with one Killing vector and there
are infinitely many solutions for cylindrical symmetry in four dimensions
\cite{line,lemo}.
For the other $\omega$ cases there are hints that
a rich black hole structure,
similar to the $\omega=0$ theory, arises \cite{chan}.

In another work \cite{lesa}
we have found the two dimensional (2D)
black holes of the corresponding generalized dilaton gravity theory.
Denoting by $\omega_3$ the Brans-Dicke parameter in 3D and by $\omega_2$
the Brans-Dicke parameter in 2D,
one finds that for causal structure
purposes  they are related by $\omega_3 = - {2\omega_2+1\over\omega_2}$.
Then by straight comparison of the Penrose diagrams one sees that the
causal structure of some of the 2D black holes
is richer than the corresponding 3D ones (e.g. $\omega_3=-\frac43$).
This is a topological effect and is due to the fact that in 3D,
if $r=0$ is a non-spacelike curve,
then it can be considered as a boundary of space while in 2D one can
continue past $r=0$ to negative values of $r$.
On the other hand,
the geodesic structure of the 3D black hole geometries
is much richer than the 2D ones,
since in 3D one can have circular, spiralling inward and
spiralling outward physical orbits.

\vskip 1cm

\noindent
{\bf Acknowledgements}

\vskip 3mm

\noindent
PMS would like to thank the {\it Departamento
de Astrof\'{\i}sica} of the {\it Observat\'orio Nacional - CNPq}
(Rio de Janeiro) for hospitality and financial support.

\newpage

\noindent
{\bf Figure Captions}

\vskip 1cm

\noindent
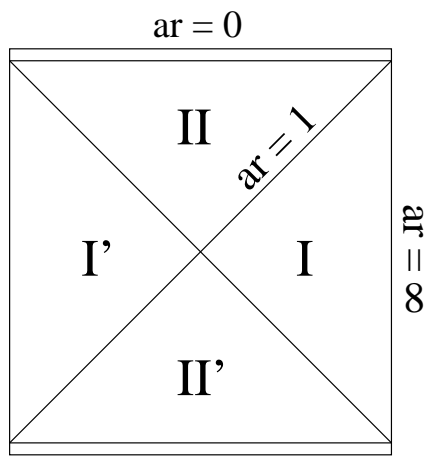
{\bf Figure 1.} Penrose diagram for the black holes with $\omega=-3$
     		and $\omega=0$, re\-pre\-sentative cases of
		$(-\infty<\omega<-\frac32)-\{-2\}$ and
		$-1<\omega<+\infty$, respectively.
		In all figures a double line represents a curvature
		singularity.

\vskip 5mm

\noindent
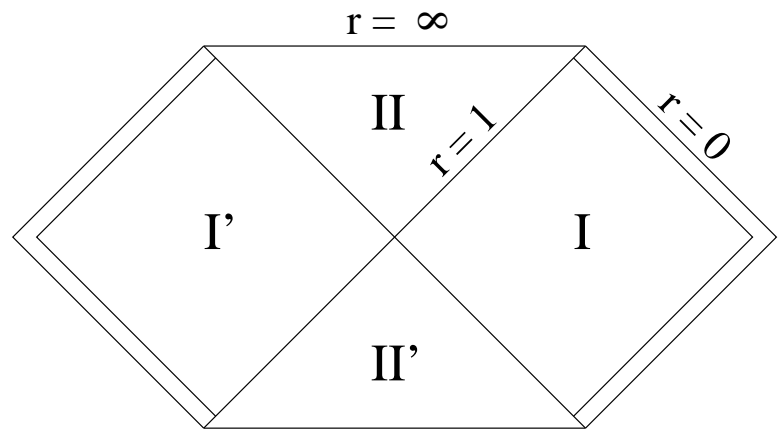
{\bf Figure 2.} Penrose diagram for $\omega=-2$.

\vskip 5mm

\noindent
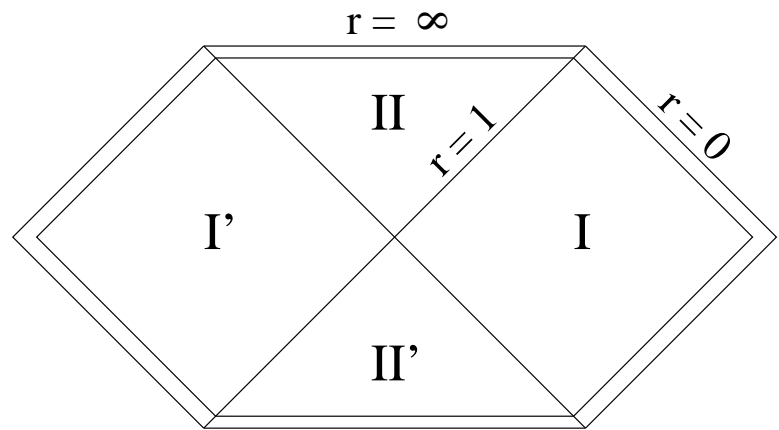
{\bf Figure 3.} Penrose diagram for $\omega=-\frac32$.

\vskip 5mm

\noindent
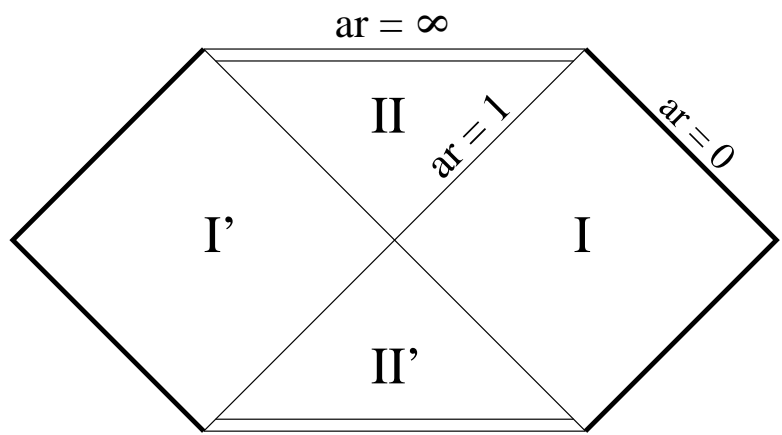
{\bf Figure 4.} Penrose diagram for the black hole with
		$\omega=-\frac43$, representative case of
		$-\frac32<\omega<-1$.
		The darker line represents a conical singularity
		at $r=0$.

\end{document}